# Propagation of Coupled Acoustic, Electromagnetic and Spin Waves in Saturated Ferromagnetoelastic Solids

Qingguo Xia[a], Jianke Du[a,*] and Jiashi Yang[b,⊥]

[a]Smart Materials and Advanced Structures Laboratory, School of Mechanical Engineering and Mechanics, Ningbo University, Ningbo, Zhejiang 315211, China
[b]Department of Mechanical and Materials Engineering, University of Nebraska-Lincoln, Lincoln, NE 68588-0526, USA

*E-mail: dujianke@nbu.edu.cn (Jianke Du)
[⊥]E-mail: jyang1@unl.edu (Jiashi Yang)

**Conflict of interest statement**: On behalf of all authors, the corresponding author states that there is no conflict of interest.

**Data availability statement**: The data that supports the findings of this study are available within the article.

**Keywords**: acoustic; electromagnetic; ferromagnetic; photon-phonon-magnon interaction

## Abstract

We study the propagation of plane waves in an unbounded body of a saturated ferromagnetoelastic solid. Tiersten's equations for small fields superposed on finite initial fields in a saturated ferromagnetoelastic material are employed, with their quasistatic magnetic field extended to dynamic electric and magnetic fields governed by Maxwell's equations for electromagnetic waves. Dispersion relations of the plane waves are obtained. The cutoff frequencies and long-wave approximation of the dispersion curves are determined. Results show that acoustic, electromagnetic and magnetic spin waves are coupled in such a material. For YIG which is a cubic crystal without piezoelectric coupling, the acoustic and electromagnetic waves are not directly coupled but they can still interact indirectly through spin waves.

## 1. Introduction

In saturated ferromagnetic solids, the magnetization vector has a fixed magnitude (saturation magnetization) and can change its direction only in a precessional motion. Below the Curie temperature, neighboring magnetization vectors align themselves in a certain direction called an easy axis of the material to form a distribution of spontaneous magnetization. A disturbance of the magnetization field propagates as spin waves with various applications [1]. Spin waves can interact with acoustic waves through magnetoelastic couplings such as piezomagnetic and magnetostrictive effects, which is called magnon-phonon interaction for which many references can be found in [2], a recent review article. Obviously, as a motion of magnetic moments, spin waves interact with electromagnetic waves directly as governed by Maxwell's equations which is referred to as photon-magnon coupling (see [3] and the references therein). Thus, in deformable ferromagnetic solids, acoustic waves and electromagnetic waves can interact indirectly through spin waves. If the material is piezoelectric, acoustic waves and electromagnetic waves are also coupled piezoelectrically. It has been reported recently [4] that the couplings among surface acoustic waves (SAW), spin waves and electromagnetic waves can be used for making electromagnetic antenna at SAW frequencies (low-frequency antenna). This has motivated our study below on the propagation of coupled acoustic, electromagnetic and spin waves (phonon-photon-magnon interaction).

## 2. Governing Equations

Consider the widely-used Yttrium Iron Garnet ($Y_3Fe_5O_{12}$) or YIG. In Gaussian units, the governing equations are [5,6]

$$\tau_{ij,i} + M_i^0 h_{j,i}^M = \rho \ddot{u}_j, \tag{1}$$

$$\nabla \cdot \mathbf{d}^M = 0, \tag{2}$$

$$\nabla \cdot \mathbf{b}^M = 0, \tag{3}$$

$$\nabla \times \mathbf{e}^M + \frac{1}{C}\frac{\partial \mathbf{b}^M}{\partial t} = 0, \tag{4}$$

$$\nabla \times \mathbf{h}^M = \frac{1}{C}\frac{\partial \mathbf{d}^M}{\partial t}, \tag{5}$$

$$\varepsilon_{ijk} M_j^0 (h_k^M - a_{lk,l} + h_k^L) + \varepsilon_{ijk} m_j H_k^0 = \frac{1}{\gamma}\dot{m}_i, \tag{6}$$

where $\tau$ is the stress tensor. $\rho$ is the mass density. $\mathbf{u}$ is the displacement vector. $\mathbf{e}^M$, $\mathbf{d}^M$, $\mathbf{b}^M$ and $\mathbf{h}^M$ are the Maxwellian electric field, electric displacement, magnetic induction and magnetic field. $C$ is the speed of light in a vacuum. $\mathbf{M}^0$ and $\mathbf{H}^0$ are the initial magnetization and initial magnetic field which are static. The initial electric and polarization fields are assumed to be zero. $\mathbf{h}^L$ is an effective local magnetic field which describes the interaction between the magnetic spin and the lattice [5]. $\mathbf{a}$ describes the exchange interaction between neighboring magnetic spins [5]. $\mathbf{m}$ is the incremental magnetization vector. $\gamma$ is the gyromagnetic ratio which is a negative number. (1) is the linear momentum equation. (2)-(5) are Maxwell's equations. (6) is the angular momentum equation of the magnetic spin. (2) and (3) are essentially implied by (4) and (5). We also have the following relationships:

$$\begin{aligned} d_i^M &= e_i^M + 4\pi p_i, \\ b_i^M &= h_i^M + 4\pi m_i - 4\pi M_i^0 u_{j,j}, \end{aligned} \tag{7}$$

where $\mathbf{p}$ is the electric polarization vector. Magnetoelectric coupling, if present, is not considered.

YIG is a cubic crystal of class (m3m). Let the spontaneous magnetization $\mathbf{M}^0$ (and $\mathbf{H}^0$) be along the $x_3$ axis. In this case $m_3=0$ because of the saturation condition $\mathbf{M}\cdot\mathbf{M}=(M^0)^2$ which implies that $\mathbf{M}^0\cdot\mathbf{m}=0$ where $\mathbf{M}=\mathbf{M}^0+\mathbf{m}$ and $\mathbf{m}$ is small. The constitutive relations are [6]

$$\begin{aligned}
\tau_1 &= \tau_{11} = c_{11}u_{1,1} + c_{12}u_{2,2} + c_{12}u_{3,3}, \\
\tau_2 &= \tau_{22} = c_{12}u_{1,1} + c_{11}u_{2,2} + c_{12}u_{3,3}, \\
\tau_3 &= \tau_{33} = c_{12}u_{1,1} + c_{12}u_{2,2} + c_{11}u_{3,3},
\end{aligned} \tag{8}$$

$$\begin{aligned}
\tau_4 &= \tau_{23} = c_{44}(u_{2,3} + u_{3,2}) + 2b_{44}M^0 m_2, \\
\tau_5 &= \tau_{31} = c_{44}(u_{1,3} + u_{3,1}) + 2b_{44}M^0 m_1, \\
\tau_6 &= \tau_{12} = c_{44}(u_{1,2} + u_{2,1}),
\end{aligned} \tag{9}$$

$$p_i = \chi^e e_i^M \quad \text{or} \quad d_i^M = \varepsilon e_i^M, \tag{10}$$

$$\begin{aligned}
h_1^L &= -\chi(M^0)^2 m_1 - 2b_{44}M^0(u_{1,3} + u_{3,1}), \\
h_2^L &= -\chi(M^0)^2 m_2 - 2b_{44}M^0(u_{2,3} + u_{3,2}), \\
h_3^L &= 0,
\end{aligned} \tag{11}$$

$$h_1^L = -\chi(M^0)^2 m_1 - 2b_{44} M^0 (u_{1,3} + u_{3,1}),$$
$$h_2^L = -\chi(M^0)^2 m_2 - 2b_{44} M^0 (u_{2,3} + u_{3,2}), \tag{12}$$
$$h_3^L = 0,$$
$$a_{ib} = -2\alpha_{11} m_{b,i}, \tag{13}$$

where

$$\rho = 5.172 \text{ g/cm}^3, \quad c_{11} = 26.9 \times 10^{11} \text{ dyn/cm}^2,$$
$$c_{12} = 10.77 \times 10^{11} \text{ dyn/cm}^2, \quad c_{44} = 7.64 \times 10^{11} \text{ dyn/cm}^2,$$
$$b_{11} - b_{12} = 1.66 \times 10^2, \quad b_{44} = 1.66 \times 10^2, \tag{14}$$
$$\chi = 3\,_4\chi_{12} - \,_4\chi_{11} = 3.36 \times 10^{-5} \text{ Oe}^{-2}, \quad \alpha_{11} = 1.87 \times 10^{-11} \text{ cm}^2,$$
$$\gamma = -1.76 \times 10^7 \text{ Oe-cm}^2/\text{dyn-sec}, \quad M^0 = 1750/4\pi \text{ G}.$$

YIG is nonpiezoelectric and nonpiezomagnetic in its natural state without any fields. Due to the spontaneous magnetization and magnetostriction, it becomes effectively piezomagnetic.

### 3. Antiplane Motion

Consider cubic crystals of class (m3m) such as YIG in Gaussian units. With the initial magnetization $\mathbf{M}^0$ and magnetic field $\mathbf{H}^0$ along the $x_3$ axis, for antiplane problems [7] with $u_1 = u_2 = 0$ and $\partial/\partial x_3 = 0$, the relevant fields are

$$u_1 = u_2 = 0, \quad u_3 = u_3(x_1, x_2, t),$$
$$e_1^M = 0, \quad e_2^M = 0, \quad e_3^M = e_3^M(x_1, x_2, t),$$
$$d_1^M = 0, \quad d_2^M = 0, \quad d_3^M = e_3^M(x_1, x_2, t), \tag{15}$$
$$b_1^M = b_1^M(x_1, x_2, t), \quad b_2^M = b_2^M(x_1, x_2, t), \quad b_3^M = 0,$$
$$h_1^M = h_1^M(x_1, x_2, t), \quad h_2^M = h_2^M(x_1, x_2, t), \quad h_3^M = 0.$$

In this case (2) is trivially satisfied. (4) and (5) reduce to

$$e_{3,2}^M + \frac{1}{C}\frac{\partial b_1^M}{\partial t} = 0, \quad -e_{3,1}^M + \frac{1}{C}\frac{\partial b_2^M}{\partial t} = 0, \tag{16}$$

$$h_{2,1}^M - h_{1,2}^M = \frac{1}{C}\frac{\partial d_3^M}{\partial t}. \tag{17}$$

We also have:

$$c_{44}(u_{3,11} + u_{3,22}) + 2b_{44} M^0 (m_{1,1} + m_{2,2}) = \rho \ddot{u}_3,$$
$$-M^0 h_2^M - 2\alpha_{11} M^0 (m_{2,11} + m_{2,22}) + \chi(M^0)^3 m_2 + 2b_{44}(M^0)^2 u_{3,2} + H^0 m_2 = \frac{1}{\gamma}\dot{m}_1,$$
$$M^0 h_1^M + 2\alpha_{11} M^0 (m_{1,11} + m_{1,22}) - \chi(M^0)^3 m_1 - 2b_{44}(M^0)^2 u_{3,1} - H^0 m_1 = \frac{1}{\gamma}\dot{m}_2, \tag{18}$$
$$h_{1,1}^M + h_{2,2}^M + 4\pi(m_{1,1} + m_{2,2}) = 0,$$

where $(18)_4$ is essentially implied by (16). Then (16), (17) and $(18)_{1-3}$ can be written as six equations for $u_3$, $e_3^M$, $h_1^M$, $h_2^M$, $m_1$ and $m_2$.

## 4. Propagation of Plane Waves

Let

$$u_3 = A_1 \exp[i(\xi x_1 - \omega t)], \quad e_3^M = A_2 \exp[i(\xi x_1 - \omega t)],$$
$$h_1^M = A_3 \exp[i(\xi x_1 - \omega t)], \quad h_2^M = A_4 \exp[i(\xi x_1 - \omega t)], \quad (19)$$
$$m_1 = A_5 \exp[i(\xi x_1 - \omega t)], \quad m_2 = A_6 \exp[i(\xi x_1 - \omega t)].$$

The substitution of (19) into (16), (17) and (18)$_{1-3}$ results in a system of six linear and homogeneous equations for $A_1$ through $A_6$. For nontrivial solutions, the determinant of the coefficient matrix has to vanish. This leads to the following equation that determines the dispersion relation of the wave:

$$-C^2 \xi^2 \left[ \alpha^2 c_{44} \xi^6 - \left( \alpha^2 \rho \omega^2 - c_{44} \alpha (2P + 4\pi) + \alpha e^2 \right) \xi^4 \right.$$
$$- \left( e^2 P - c_{44} P(P + 4\pi) + \omega^2 \left( \frac{c_{44}}{\gamma^2 (M^0)^2} + \rho \alpha (2P + 4\pi) \right) \right) \xi^2$$
$$\left. - \rho P(P + 4\pi) \omega^2 + \frac{\rho}{\gamma^2 (M^0)^2} \omega^4 \right]$$
$$+ \varepsilon \omega^2 \left[ c_{44} \alpha^2 \xi^6 - \left( \alpha^2 \rho \omega^2 - 2 c_{44} \alpha (P + 4\pi) + \alpha e^2 \right) \xi^4 \right. \quad (20)$$
$$- \left( e^2 (P + 4\pi) - c_{44} (P + 4\pi)^2 + \omega^2 \left( \frac{c_{44}}{\gamma^2 (M^0)^2} + 2 \rho \alpha (P + 4\pi) \right) \right) \xi^2$$
$$\left. - \rho (P + 4\pi)^2 \omega^2 + \frac{\rho}{\gamma^2 (M^0)^2} \omega^4 \right] = 0,$$

where

$$e = 2 b_{44} M^0, \quad \alpha = 2 \alpha_{11}, \quad P = H^0 / M^0 + K, \quad K = \chi (M^0)^2. \quad (21)$$

In the special case of $b_{44}=0$, the magnetoelastic coupling disappears and (20) reduces to the product of two factors. One is for uncoupled acoustic waves:

$$\rho \omega^2 - c_{44} \xi^2 = 0. \quad (22)$$

The other is for coupled electromagnetic and spin waves:

$$-C^2 \xi^2 \left[ \frac{\omega^2}{\gamma^2 (M^0)^2} - (\alpha \xi^2 + P)(\alpha \xi^2 + P + 4\pi) \right] + \varepsilon \omega^2 \left[ \frac{\omega^2}{\gamma^2 (M^0)^2} - (\alpha \xi^2 + P + 4\pi)^2 \right] = 0. \quad (23)$$

When $C \to \infty$, (23) reduces to the following dispersion relation for uncoupled spin waves:

$$\frac{\omega^2}{\gamma^2 (M^0)^2} = \alpha^2 \xi^4 + \alpha (2P + 4\pi) \xi^2 + P(P + 4\pi). \quad (24)$$

When $\alpha=0$ and $\gamma \to \infty$, (23) reduces to the following dispersion relation for uncoupled electromagnetic waves

$$\frac{\omega^2}{\xi^2} = \frac{C^2 P}{\varepsilon (P + 4\pi)}. \quad (25)$$

When $\varepsilon=1$ and $M^0 \to 0$, we have $P \to \infty$ and (25) reduces to $\omega/\xi=C$ for electromagnetic waves in a vacuum.

When $H^0=1500$ Oe and $\varepsilon=14$, the dispersion relations of the uncoupled waves in (22), (24) and (25) are shown in Fig. 1 in logarithmic scales for both the coordinate and the abscissa. The acoustic and electromagnetic waves are represented by straight lines and are nondispersive, with the electromagnetic waves at higher frequencies. The spin wave is represented by a curve. Each straight line intersects with the curve at two points.

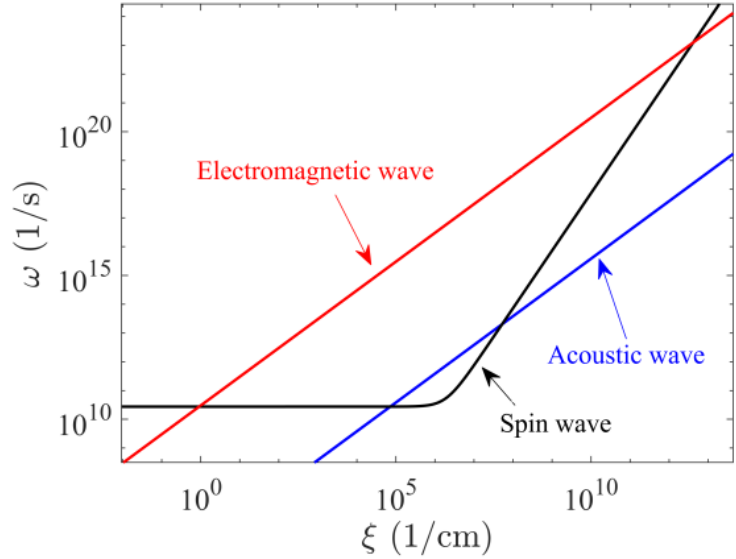

Fig. 1. Uncoupled acoustic, electromagnetic and spin waves.

When $H^0$=1500 Oe and $\varepsilon$=14, the dispersion relations of the coupled waves determined by (20) are shown in Fig. 2. Near B and C there are strong couplings between acoustic and spin waves. Near A and D there are strong couplings between electromagnetic and spin waves. Since the material is nonpiezoelectric, there is no direct coupling between acoustic and electromagnetic waves.

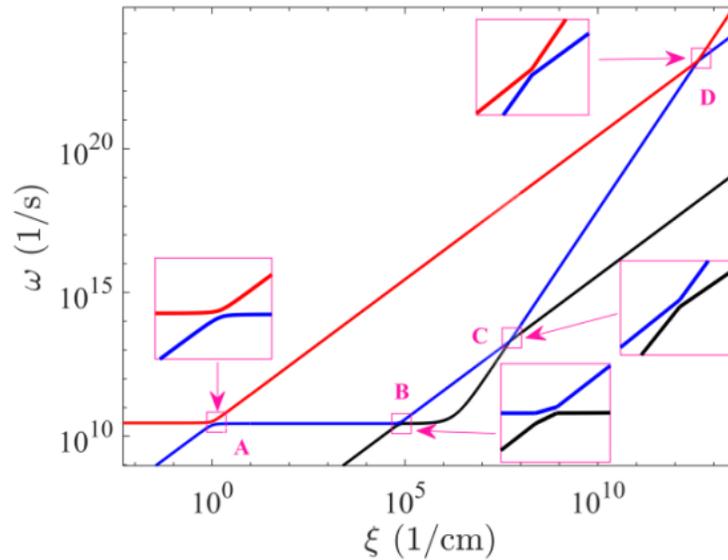

Fig. 2. Dispersion curves of coupled waves when $H^0$=1500 Oe and $\varepsilon$=14.

$H^0$ is an independent parameter. If it is varied a little, its effects on the dispersion curves are shown in Fig. 3. The spin waves are sensitive to $H^0$ but the two other waves are not, which is reasonable.

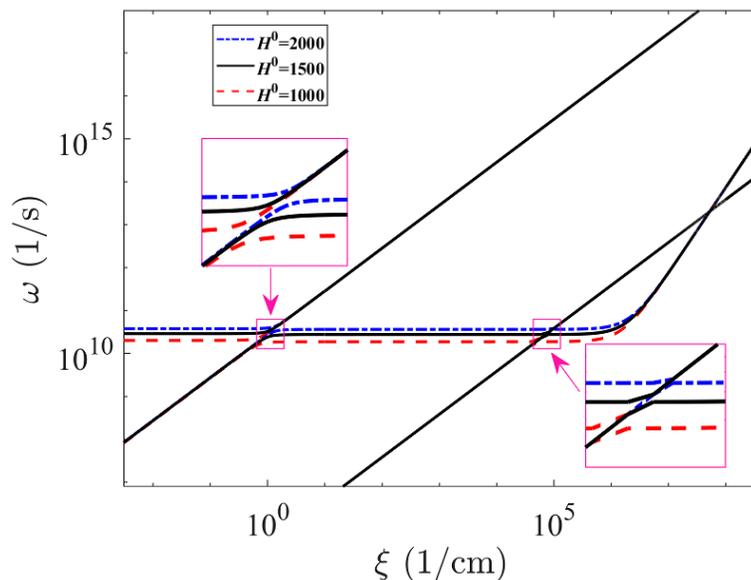

Fig. 3. Effects of $H^0$ (in Oe) on dispersion relations of coupled waves. $\varepsilon=14$.

The numerical value of $\varepsilon$ for YIG in the literature ranges from 14 to 18. The effect of slightly different values of $\varepsilon$ on the dispersion curves of the coupled waves is shown in Fig. 4 where $\varepsilon$ is denoted by $\varepsilon_r$ which mainly affects the electromagnetic waves as expected.

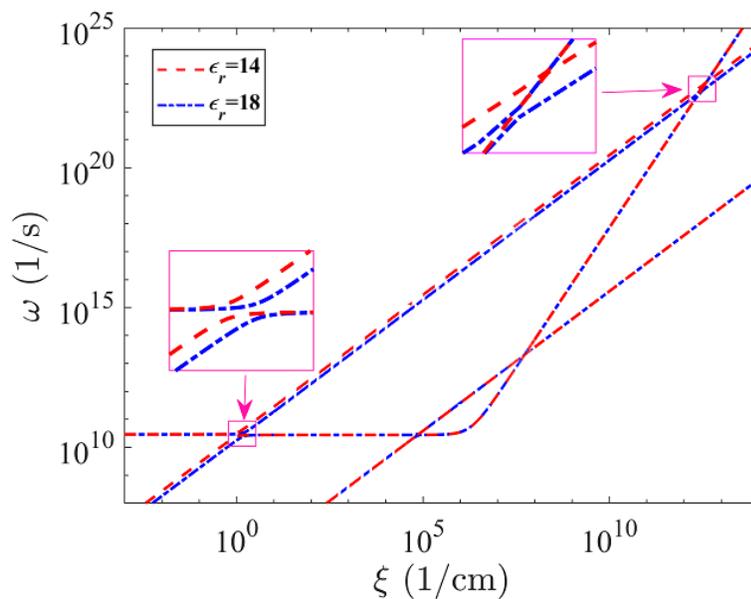

Fig. 4. Effects of $\varepsilon_r=\varepsilon$ on dispersion curves of coupled waves. $H^0=1500$ Oe.

## 5. Conclusions

In saturated ferromagnetoelastic solids such as YIG, acoustic, electromagnetic and magnetic spin waves are coupled. Thus it is possible to manipulate one wave by another or design transducers using the couplings of these waves. This offers more possibilities for new devices. At present, the literature on the three-wave coupling of photons, phonons and magnons are limited, with an absence of the mechanics community. Since the equations of elasticity represent a major part of the coupled theory for these three waves, mechanics researchers can play an important role in this interdisciplinary area.

## Acknowledgment

This work was supported by the National Natural Science Foundation of China (Nos. 12072167 and 11972199), the Zhejiang Provincial Natural Science Foundation of China (No. LR12A02001), and the K. C. Wong Magana Fund through Ningbo University.